\documentclass[twocolumn,secnumroman,amssymb, amsmath, nofootinbib,tightenlines,
nobibnotes, aps, prb, showpacs]{revtex4}

\usepackage{graphicx}
\usepackage{dcolumn}
\usepackage{amsmath}
\usepackage{here}%

\begin{document}

\preprint{HEP/123-QED}

\title{Superconductivity under pressure in $R$FeAsO$_{\rm 1-x}$F$_{\rm x}$ ($R$=La, Ce$-$Sm) \\ by dc magnetization measurements} 

\author{Kiyotaka Miyoshi, Eiko Kojima, Saki Ogawa, Yuta Shimojo, and Jun Takeuchi}
\affiliation{%
Department of Material Science, Shimane University, Matsue 690-8504, Japan
}%





\date{\today}

\begin{abstract}
Superconducting transition under pressure ($P$) has been investigated 
for optimum-doped $R$FeAsO$_{\rm 1-x}$F$_{\rm x}$ ($R$=La, Ce$-$Sm) 
by dc magnetization measurements. For $R$=La, $T_{\rm c}$ is found to be pressure independent 
up to $P$$\sim$3.0 GPa and then shows a monotonic decrease with increasing pressure. 
The plateau width decreases for the system with smaller lattice constants, and shrinks to almost zero for $R$=Sm. 
From the $T_{\rm c}$($P$) data, we construct the $T_{\rm c}$ evolution map 
on the height of the As atom from the Fe-plane $h_{\rm As}$ versus lattice constant $a$ or $c$ plane. 
It is shown that the characteristic $T_{\rm c}$ variations under pressure for all compounds and 
the rapid decrease in $T_{\rm c}$ induced by changing $R$-element from Sm to La 
can be described by a common $T_{\rm c}$($h_{\rm As}$, $a$ or $c$) surface, suggesting that 
$T_{\rm c}$ is determined by $h_{\rm As}$ and lattice constant. 
It is also suggested that there exists an upper limit of the lattice constant 
$a_{\rm ulm}$ (or $c_{\rm ulm}$), above which $dT_{\rm c}$$/$$da$ (or $dT_{\rm c}$$/$$dc$) changes 
the sign from positive to negative.

\end{abstract}

\maketitle


\section{\label{sec:level1}Introduction}

Since the discovery of superconductivity in LaFeAsO$_{\rm 1-x}$F$_{\rm x}$ ($T_{\rm c}$=26 K),\cite{kamihara} 
a great deal of progress has been made in exploring superconductivity in the related compounds, 
leading to a rich variety of iron-pnictide superconductors, 
such as Ba$_{\rm 1-x}$K$_{\rm x}$Fe$_2$As$_2$ ($T_{\rm c}$=38 K),\cite{rotter,sasmal} 
Li$_{\rm 1-x}$FeAs ($T_{\rm c}$=18 K),\cite{tapp,wang} FeSe ($T_{\rm c}$=8 K),\cite{hsu} 
and K$_{0.8}$Fe$_2$Se$_2$ ($T_{\rm c}$=32 K),\cite{guo} 
in addition to $R$FeAsO$_{\rm 1-x}$F$_{\rm x}$ ($R$=lanthanoid) ($T_{\rm c}$=26$-$53 K), where 
the superconductivity is developed in iron-pnictide (Fe-Pn) layers consisting of edge-sharing FePn$_4$ tetrahedron. 
Indeed, there exists an intimate correlation between the crystal structure and 
$T_{\rm c}$ in the iron-pnictide family, 
as demonstrated in $R$FeAsO$_{\rm 1-x}$F$_{\rm x}$ that $T_{\rm c}$ becomes maximum when FeAs$_4$ forms 
a regular tetrahedron, i.e., the As-Fe-As bond angle is $\sim$109.5$^{\circ}$.\cite{lee}  
Also, the pnictogen height $h_{\rm Pn}$ measured from the Fe plane is known to be an important structural parameter, 
which is originally introduced to act as a switch between high-$T_{\rm c}$ state with nodeless paring to 
low-$T_{\rm c}$ state with nodal paring.\cite{kuroki}
There are some attempts to plot $T_{\rm c}$ versus $h_{\rm Pn}$ for typical iron pnictide superconductors, 
suggesting that the data collapse to a universal curve.\cite{mizuguchi,okabe} 

Application of pressure can induce the modification of the crystal structure leading to the 
change in $T_{\rm c}$. Thus, to obtain the intrinsic relation between $T_{\rm c}$ and structural parameters 
under pressure is of significant importance, providing valuable information to elucidate 
the mechanism of superconductivity. One of the attractive subjects for studying the pressure effect is 
$R$FeAsO$_{\rm 1-x}$F$_{\rm x}$, where $T_{\rm c}$ systematically changes by replacing $R$ element,
so that we can extract more information by comparing the variations of $T_{\rm c}$ induced by 
physical and chemical pressure to specify which parameters play a crucial role for 
the superconductivity. 

The pressure effect for $R$FeAsO$_{\rm 1-x}$F$_{\rm x}$ has been investigated by many researchers
through the measurements of electrical resistivity ($\rho$),\cite{takahashi,okada,takeshita,lorenz,zocco, 
yi,yi2,garbarino,kawakami,sun,nakano,yamaoka,zocco2,garbarino2}  
ac and dc magnetization,\cite{lu,takabayashi,bi,dong,khasanov} X-ray diffraction,\cite{garbarino,kumai,garbarino2} 
NMR,\cite{tatsumi,nakano} and others.\cite{kawakami,khasanov}
For LaFeAsO$_{\rm 1-x}$F$_{\rm x}$ with $x$=0.11, it has been suggested that 
$T_{\rm c}$ increases with increasing pressure $P$ and 
shows a maximum of 43 K at $P$$\sim$4 GPa through the $\rho$($T$) measurements by 
determining $T_{\rm c}$ from the onset of resistive drop.\cite{takahashi,okada}
In the study, if $T_{\rm c}$ is determined by the zero resistive temperature, which is fairly lower than the onset, 
the $T_{\rm c}$$-$$P$ curve would be a quite different one showing a gradual increase from $\sim$23 K to $\sim$28 K 
for 0$\leq$$P$$\leq$3 GPa.\cite{takahashi}   
Since the resistivity drop usually occurs over a wide temperature range when the polycrystalline sample is used, 
it is extremely difficult to determine $T_{\rm c}$ precisely only by the $\rho$($T$) data. 
Magnetization data, by which $T_{\rm c}$ is determined solely from diamagnetic onset, should be used together with $\rho$($T$) 
data. However, to our knowledge, the magnetic measurements for $R$FeAsO$_{\rm 1-x}$F$_{\rm x}$ have been 
limited to the low pressure range below $\sim$1 GPa.\cite{lu,takabayashi,bi,dong,khasanov}  
 
In the present work, we have performed DC magnetization measurements for optimum-doped $R$FeAsO$_{\rm 1-x}$F$_{\rm x}$ 
($R$=La, Ce$-$Sm) under pressure using a diamond anvil cell (DAC) to establish the $T_{\rm c}$$-$$P$ relation. 
Our dc magnetic measurement using DAC is a powerful technique to determine $T_{\rm c}$ under pressure, 
and has been successfully applied to other superconductors.\cite{miyoshi1, miyoshi2} 
In this paper, it is found that the $T_{\rm c}$$-$$P$ curve for $R$=La 
is pressure independent up to $\sim$3.0 GPa and decreases monotonically above 3 GPa. 
The plateau width becomes narrower for the system with smaller lattice constants and 
almost shrinks to zero for $R$=Sm. 
These behaviors, and also the $T_{\rm c}$ variation induced by the chemical substitution, 
can be explained by considering the effect of the As height 
$h_{\rm As}$ and the lattice constant on $T_{\rm c}$ together with an upper limit of the 
lattice constant, across which $dT_{\rm c}$$/$$da$ (or $dT_{\rm c}$$/$$dc$) changes the sign. 

\section{Experimental}
The polycrystalline samples of optimum-doped $R$FeAsO$_{\rm 1-x}$F$_{\rm x}$ ($R$=La, Ce$-$Sm) 
were synthesized by a solid-state reaction technique\cite{martinelli} heating the 
pelletized mixtures of $R$As, Fe, Fe$_2$O$_3$ and FeF$_2$ powders at 1100 $^{\circ}{\rm C}$ for 20 h. 
The pellet was wrapped with Ta foil and put in an evacuated silica tube. 
$R$As was prepared by reacting $R$ and As powders in an evacuated silica tube 
at 250-1000 $^{\circ}{\rm C}$, as described in the 
literature.\cite{hiscocks} All operations for the synthesis were carried out in an Ar atmosphere. 
The samples were confirmed to be a single phase by X-ray diffraction. 
For the optimum doping, our specimens were prepared with nominal fluorine content $x$ ranging from 0.1 to 0.2. 
$T_{\rm c}$ of our specimens determined by 
the diamagnetic onset at ambient pressure was $\sim$27 K (R=La), $\sim$38 K (Ce), 
$\sim$43 K (Pr), $\sim$48 K (Nd), and $\sim$53 K (Sm), consistent with the previous reports.\cite{ghoshray,ren,tarantini,liu} 
For the magnetic measurements under high pressure, a miniature DAC with an outer 
diameter of 8 mm was used to generate high pressure and 
combined with a sample rod of a commercial SQUID magnetometer. 
The details of the DAC are given elsewhere.\cite{mito}
The sample was loaded into the gasket hole together 
with a small piece of high-purity lead (Pb) to realize the $in$ $situ$ observation of pressure
by determining the pressure from the $T_{\rm c}$ shift of Pb. 
Magnetization data for the small amounts of $R$FeAsO$_{\rm 1-x}$F$_{\rm x}$ and Pb were obtained by subtracting the 
magnetic contribution of DAC measured in an empty run from the total magnetization data. 
Daphne oil 7373 was used as a pressure transmitting medium. 
For the measurements in the high pressure regime, liquid Ar was also used
to apply hydrostatic pressure. 

\section{Results}
\begin{figure}[h]
\includegraphics[width=8.5cm]{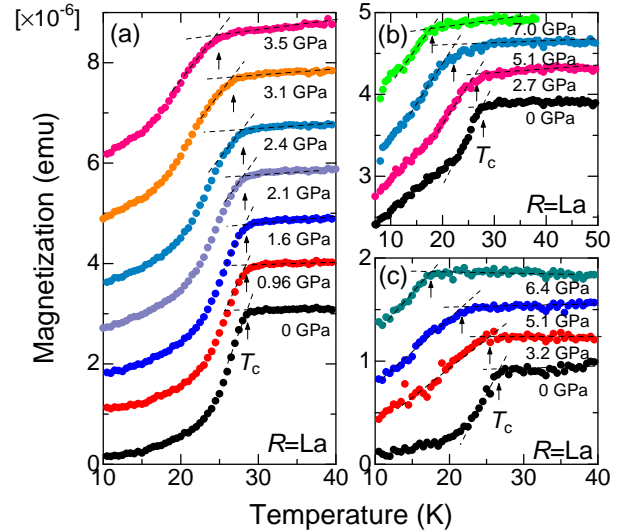}
\caption{(Color online) Temperature dependence of zero-field-cooled dc magnetization 
measured with a magnetic field of $H$=20 Oe under various pressures 
up to 3.5 GPa (a), 7.0 GPa (b), and 7.5 GPa (c) 
for LaFeAsO$_{\rm 1-x}$F$_{\rm x}$. 
The data are intentionally shifted along longitudinal axis for clarity. 
The data measured using liquid Ar as a pressure transmitting medium are shown in (c). 
}
\label{autonum}
\end{figure}
\begin{figure}[h]
\includegraphics[width=8.5cm]{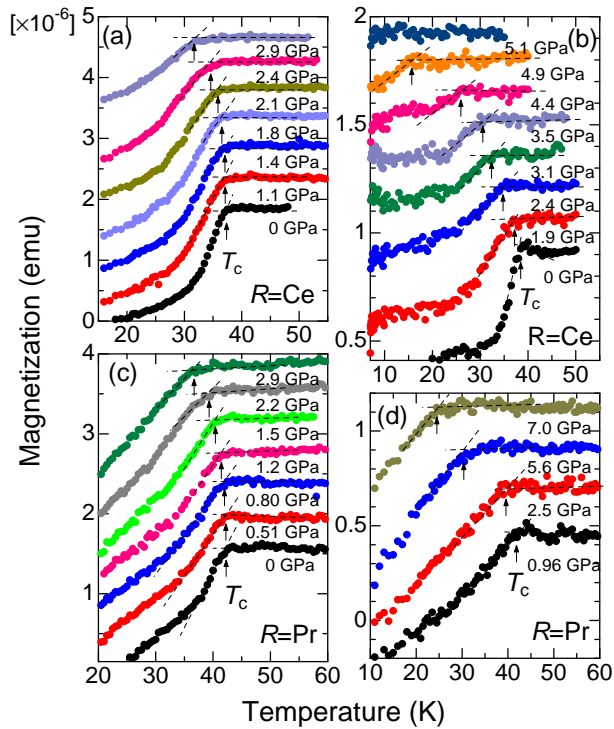}
\caption{(Color online) Temperature dependence of zero-field-cooled dc magnetization measured 
with a magnetic field of $H$=20 Oe under various pressures up to 2.9 GPa (a) and 
5.1 GPa (b) for CeFeAsO$_{\rm 1-x}$F$_{\rm x}$, and 2.9 GPa (c) and 7.0 GPa (d) 
for PrFeAsO$_{\rm 1-x}$F$_{\rm x}$. 
The data are intentionally shifted along the longitudinal axis for clarity. 
The data measured using liquid Ar as a pressure transmitting medium are shown in (b) and (d). 
}
\label{autonum}
\end{figure}
In this section, we show typical zero-field cooled dc magnetization ($M$) versus temperature ($T$) curves
for $R$=La and Ce-Sm under various pressures. $T_{\rm c}$ was determined 
by the onset temperature of diamagnetic response in 
the $M$$-$$T$ curve. The onset temperature was estimated by extrapolating the initial slope of the $M$$-$$T$ curve 
just below $T_{\rm c}$ to the normal state magnetization. 
Figure 1(a) shows the $M$$-$$T$ curves at low pressures below 3.5 GPa for $R$=La. 
At ambient pressure, the $M$$-$$T$ curve exhibits a sudden decrease at $\sim$28 K, indicating 
the onset of the diamagnetic response accompanied by the superconducting transition at $T_{\rm c}$$\sim$28 K. 
For 0$\leq$$P$$\leq$2.4 GPa, the onset temperature does not appear to change, 
indicating that $T_{\rm c}$ is pressure independent. Above 3.1 GPa, $T_{\rm c}$ is found to gradually 
shift to lower temperature side. In Figs. 1(b)-1(c), we show the typical $M$$-$$T$ curves at 
higher pressures. As increasing pressure, $T_{\rm c}$ decreases slowly and reaches $\sim$18 K at $P$=7.0 GPa. 
Next, we show the typical $M$$-$$T$ curves for $R$=Ce and Pr in Figs. 2(a)-2(d). For $R$=Ce, a sharp diamagnetic 
response is seen at ambient pressure below $T_{\rm c}$$\sim$38 K, but $T_{\rm c}$ is unchanged 
at least up to $P$=1.4 GPa, similar to the behavior seen for $R$=La, 
and then decreased gradually above $P$=1.8 GPa, as seen in Fig. 2(a). 
In Fig. 2(b), $T_{\rm c}$ is found to be 25 K at $P$=4.4 GPa, above which $T_{\rm c}$ is 
however rapidly decreased with increasing pressure and diamagnetic response was not observed at $P$=5.1 GPa above 5 K, 
suggesting that the superconductivity is suppressed under pressure above 5 GPa. 
Disappearance of superconductivity under pressure for $R$=Ce has been previously reported 
in earlier studies,\cite{takeshita2,sun,yamaoka} where the origin is discussed in terms of 
the valence transition of Ce ion and the competition between superconductivity and Kondo screening state. 

In Fig. 2(c), both of the $M$$-$$T$ curve for $R$=Pr at ambient pressure and at $P$=0.51 GPa indicates 
a superconducting transition at $T_{\rm c}$$\sim$43 K. Above $P$=0.80 GPa, $T_{\rm c}$ begins to shift toward 
lower temperature side, indicating that $T_{\rm c}$ is nearly constant in the 
pressure range below 0.8 GPa, which is lower than that observed for $R$=La and Ce.  
With further pressure increase, 
$T_{\rm c}$ for $R$=Pr shows a monotonic decrease and reaches $\sim$25 K at 7.0 GPa, as seen in Fig. 2(d). 
Figures 3(a)-3(d) show the $M$$-$$T$ curves for $R$=Nd and Sm. 
$T_{\rm c}$ for $R$=Nd appears to be $\sim$48 K at $P$=0.39 GPa in Fig. 3(a). A remarkable shift of 
$T_{\rm c}$ is not observed up to $P$=0.74 GPa but further application of pressure suppresses $T_{\rm c}$ 
down to 26 K at $P$=7.7 GPa as seen in Fig. 3(b). The pressure range in which $T_{\rm c}$ is pressure independent 
is similar to that for $R$=Pr. On the other hand, it is found that the diamagnetic onset 
in the $M$$-$$T$ curve for $R$=Sm at 0.36 GPa 
is slightly lowered from that at ambient pressure ($\sim$53 K), 
and then monotonously decreased down to $\sim$48 K at 2.5 GPa and $\sim$29 K at 7.7 GPa
by the application of pressure, as seen in Figs. 3(c) and 3(d). 
$T_{\rm c}$ for $R$=Sm is immediately decreased without delay even in the low pressure region. 
\begin{figure}[h]
\includegraphics[width=8.5cm]{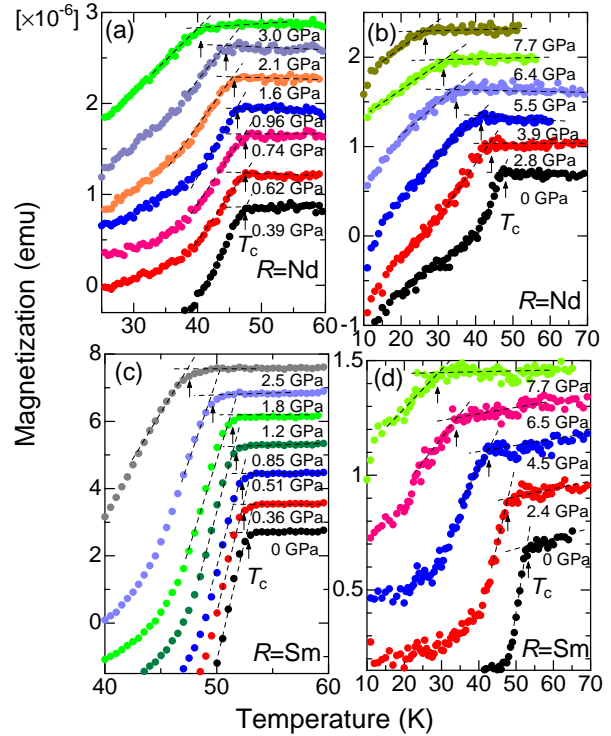}
\caption{(Color online) Temperature dependence of zero-field-cooled dc magnetization measured with a magnetic field of $H$=20 Oe 
under various pressures up to 3.0 GPa (a) and 
7.7 GPa (b) for NdFeAsO$_{\rm 1-x}$F$_{\rm x}$, and up to 2.5 GPa (c) and 7.7 GPa (d) 
for SmFeAsO$_{\rm 1-x}$F$_{\rm x}$. 
The data are intentionally shifted along the longitudinal axis for clarity. 
The data measured using liquid Ar as a pressure transmitting medium are shown in (d). }
\label{autonum}
\end{figure}

Next, we show plots of $T_{\rm c}$ versus $P$ data for $R$=La and Ce-Sm in Fig. 4. 
In the figure, a characteristic plateau is seen in the $T_{\rm c}$$-$$P$ curve for $R$=La in the 
pressure range below $P_{\rm w}$$\sim$3.0 GPa. For the system with smaller lattice constants, 
the plateau width in the $T_{\rm c}$$-$$P$ curve is 
found to be narrower, i.e., $P_{\rm w}$$\sim$1.5 GPa for $R$=Ce, $P_{\rm w}$$\sim$0.5$-$1.0 GPa for $R$=Pr and Nd, 
and $P_{\rm w}$$<$0.5 GPa for $R$=Sm. For $P$$\geq$$P_{\rm w}$, the $T_{\rm c}$$-$$P$ curve exhibits a 
monotonic decrease except for $R$=Ce. $T_{\rm c}$ for $R$=Ce decreases rapidly especially above $P$=4 GPa 
dropping toward $T_{\rm c}$=0 at $P$=5 GPa. The decreasing rates $d$$T_{\rm c}$$/$$dP$ for $R$=Pr$-$Sm are similar to 
each other yielding $-$3$-$4 K$/$GPa. 
In the figure, it is also found that the $T_{\rm c}$$-$$P$ relation does not depend on whether the pressure transmitting 
medium is liquid Ar or Daphne oil 7373. 
\begin{figure}[h]
\includegraphics[width=7.5cm]{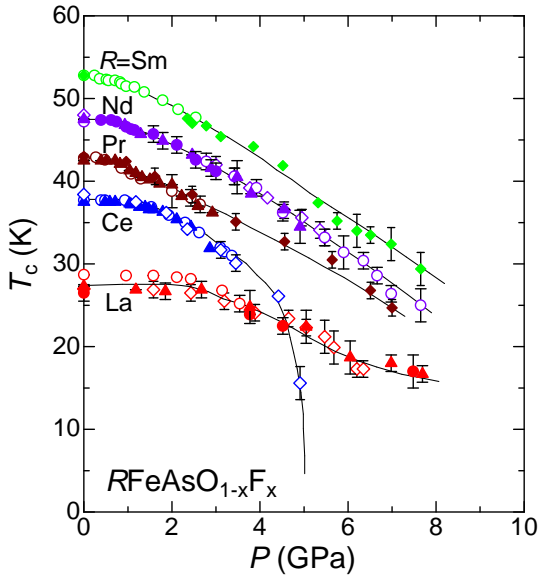}
\caption{(Color online) Pressure variations of critical temperature $T_{\rm c}$ for $R$FeAsO$_{\rm 1-x}$F$_{\rm x}$ ($R$=La, Ce-Sm). 
The solid lines are guides for the eyes. 
The data sets measured in different runs are plotted by different symbols. 
Open and closed diamonds (other symbols) 
indicate the data obtained by using liquid Ar (Daphne oil 7373) as the pressure transmitting medium. 
}
\label{autonum}
\end{figure}

\section{Discussion}
\subsection{Intrinsic $T_{\rm c}$$-$$P$ relation}
One may note that the plateau behavior in the $T_{\rm c}$$-$$P$ curve for $R$=La seen in Fig. 4 
is inconsistent with the $T_{\rm c}$$-$$P$ relation reported by Takahashi $et$ $al$., 
which is determined from the onset of resistive drop 
for the specimen with $F$-content $x$=0.11, showing a maximum of 
$T_{\rm c}$$\sim$43 K at $P$$\sim$4 GPa.\cite{takahashi}
The discrepancy originates from not only the difference in the definition of $T_{\rm c}$ 
but also the difference in the doping level of the specimens, because their specimen with $x$=0.11 
shows a relatively low diamagnetic 
onset temperature of $\sim$22 K at ambient pressure.\cite{takahashi}
On the other hand, $T_{\rm c}$ for their specimen with $x$=0.05 
is $\sim$23 K,\cite{takahashi} 
and $T_{\rm c}$ for $x$=0.08 is found to be $\sim$28 K,\cite{nakano}  
indicating that their specimen with $x$=0.11 (0.08) is overdoped (optimum-doped). 
For $x$=0.08, it has been shown that 
the zero resistive temperature in the $\rho$($T$) curve ($\sim$28 K) is unchanged at least up to  
2.60 GPa.\cite{nakano} The behavior is the same with that observed for $R$=La, as  
shown in the $T_{\rm c}$$-$$P$ curve in Fig. 4, 
indicating that the zero resistive temperature and the diamagnetic onset are identical to 
each other, and both of them can be reliable markers of $T_{\rm c}$ for $R$FeAsO$_{\rm 1-x}$F$_{\rm x}$. 
We therefore suggest that the behavior showing a plateau for 0$\leq$$P$$\leq$3 GPa 
is the intrinsic $T_{\rm c}$$-$$P$ relation for optimally doped $R$=La. 

Pressure dependence of $T_{\rm c}$ has been widely investigated by the $\rho$($T$) and $M$($T$) 
measurements for $R$=Sm,\cite{lorenz,yi,takabayashi,dong,garbarino2} 
Nd,\cite{takeshita}Pr,\cite{dong} and Ce,\cite{sun,yamaoka,zocco} 
and negative pressure coefficients of $T_{\rm c}$ have been found in these studies. 
In particular, the investigations in high pressure range have been done by the $\rho$($T$) measurements adopting the 
onset temperature of resistive drop as $T_{\rm c}$ for $R$=Sm and Nd, 
resulting in $d$$T_{\rm c}$$/$$dP$$\sim$$-$3.0 K/GPa.\cite{garbarino2,takeshita} 
The value is similar to that obtained for $R$=Sm-Pr in the present study.    
For $R$=Ce, the $T_{\rm c}$$-$$P$ relation has been also investigated through the $\rho$($T$) measurements, 
suggesting that the superconductivity disappears at $P$=4.5$-$5 GPa,\cite{yamaoka} similar to the result shown in Fig. 4. 
\begin{figure}[h]
\includegraphics[width=7.5cm]{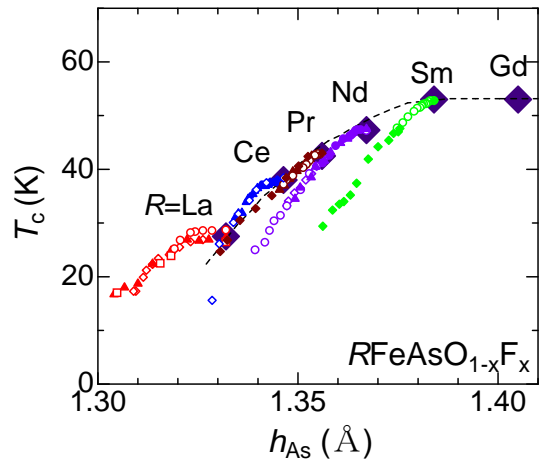}
\caption{(Color online) Plots of $T_{\rm c}$ versus height of As layer from Fe layer $h_{\rm As}$ 
for $R$FeAsO$_{\rm 1-x}$F$_{\rm x}$ ($R$=La, Ce-Sm). The broken line interpolates the data 
at ambient pressure, which are displayed by large closed diamonds, 
and corresponds to the variation for the chemical substitution for $R$ site. 
Small symbols represent the data obtained from the physical pressure dependence of $T_{\rm c}$. 
The $T_{\rm c}$($P$) data are transformed to $T_{\rm c}$($h_{\rm As}$) data assuming a linear relation with a  
coefficient $dh_{\rm As}$$/$$dP$$\sim$$-$3.6$\times$10$^{-3}$ \AA \ GPa$^{-1}$ 
extracted from the pressure variations of structural parameters in the literature.\cite{garbarino2}  
The values of $h_{\rm As}$ at ambient pressure can be obtained from the literatures.\cite{okabe,lee,nomura,zhao} 
}
\label{autonum}
\end{figure}

\subsection{Plots of $T_{\rm c}$ versus $h_{\rm As}$}
Pnictogen height from the Fe layer $h_{\rm Pn}$ is 
considered to be a key factor to determine $T_{\rm c}$ in an iron-pnictide superconductor. 
For FeSe superconductor, it has been found that the pressure variations of $T_{\rm c}$ and 
Se-height $h_{\rm Se}$ are qualitatively analogous to each other.\cite{okabe} 
Also for SmFeAsO$_{\rm 1-x}$F$_{\rm x}$, an attempt to compare the pressure variations 
of $T_{\rm c}$ and $h_{\rm As}$ has been made, confirming that both of them shows 
a monotonous decrease under pressure above 1 GPa.\cite{garbarino2} 
As shown theoretically, the increase of $h_{\rm As}$ leads to the appearance of $\gamma$ Fermi surface, 
resulting in the enhancement of $T_{\rm c}$ (i.e., $dT_{\rm c}$$/$$dh_{\rm As}$$>$0) in $R$FeAsO system.\cite{kuroki}
In order to examine the importance of $h_{\rm As}$, 
it is interesting to compare the $T_{\rm c}$ versus $h_{\rm As}$ data derived from 
the structural modulation originating from the physical compression and 
the chemical substitution for $R$-site. 
If $h_{\rm As}$ is the only structural parameter to determine $T_{\rm c}$, 
the $T_{\rm c}$-$h_{\rm As}$ data under physical and chemical pressure would coincide with each other. 
To obtain $T_{\rm c}$($h_{\rm As}$) data, we transform the $T_{\rm c}$($P$) data 
assuming a linear relation between $h_{\rm As}$ and $P$ with a coefficient 
$dh_{\rm As}$$/$$dP$$\sim$$-$3.6$\times$10$^{-3}$ \AA \ GPa$^{-1}$, 
which is extracted from the pressure variations of structural parameters for SmFeAsO$_{\rm 1-x}$F$_{\rm x}$ 
interpolating the data for 0.5$\leq$$P$$\leq$20 GPa in the literature.\cite{garbarino2} 
The value of $dh_{\rm As}$$/$$dP$ well agrees with that obtained from the 
first principle calculations of the pressure effect on LaFeAsO, that is 
$dh_{\rm As}$$/$$dP$$\sim$$-$3.5$\times$10$^{-3}$ \AA \ GPa$^{-1}$.\cite{nakamura} 
\begin{figure}[!h]
\includegraphics[width=6.5cm]{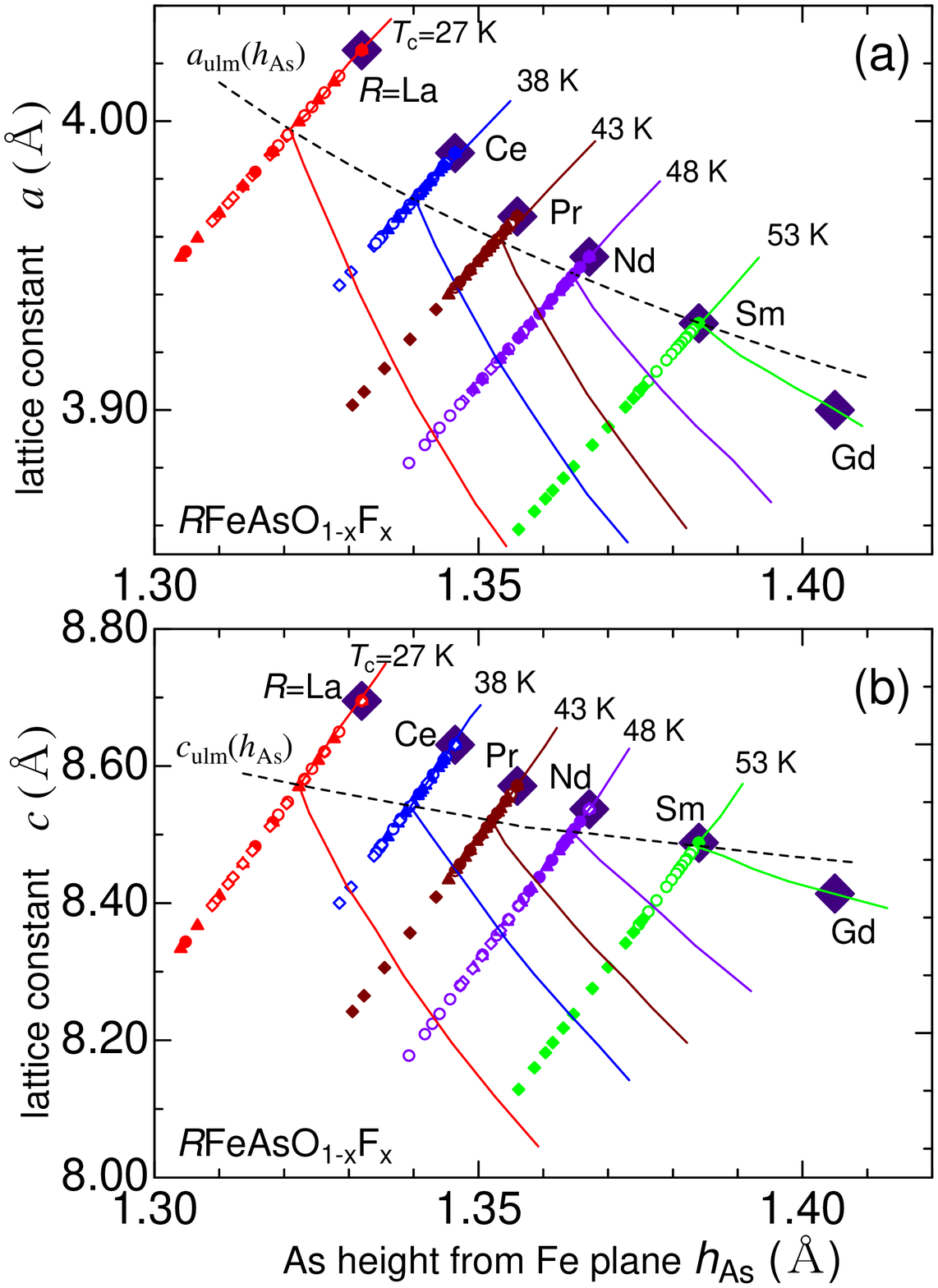}
\\
\includegraphics[width=7.5cm]{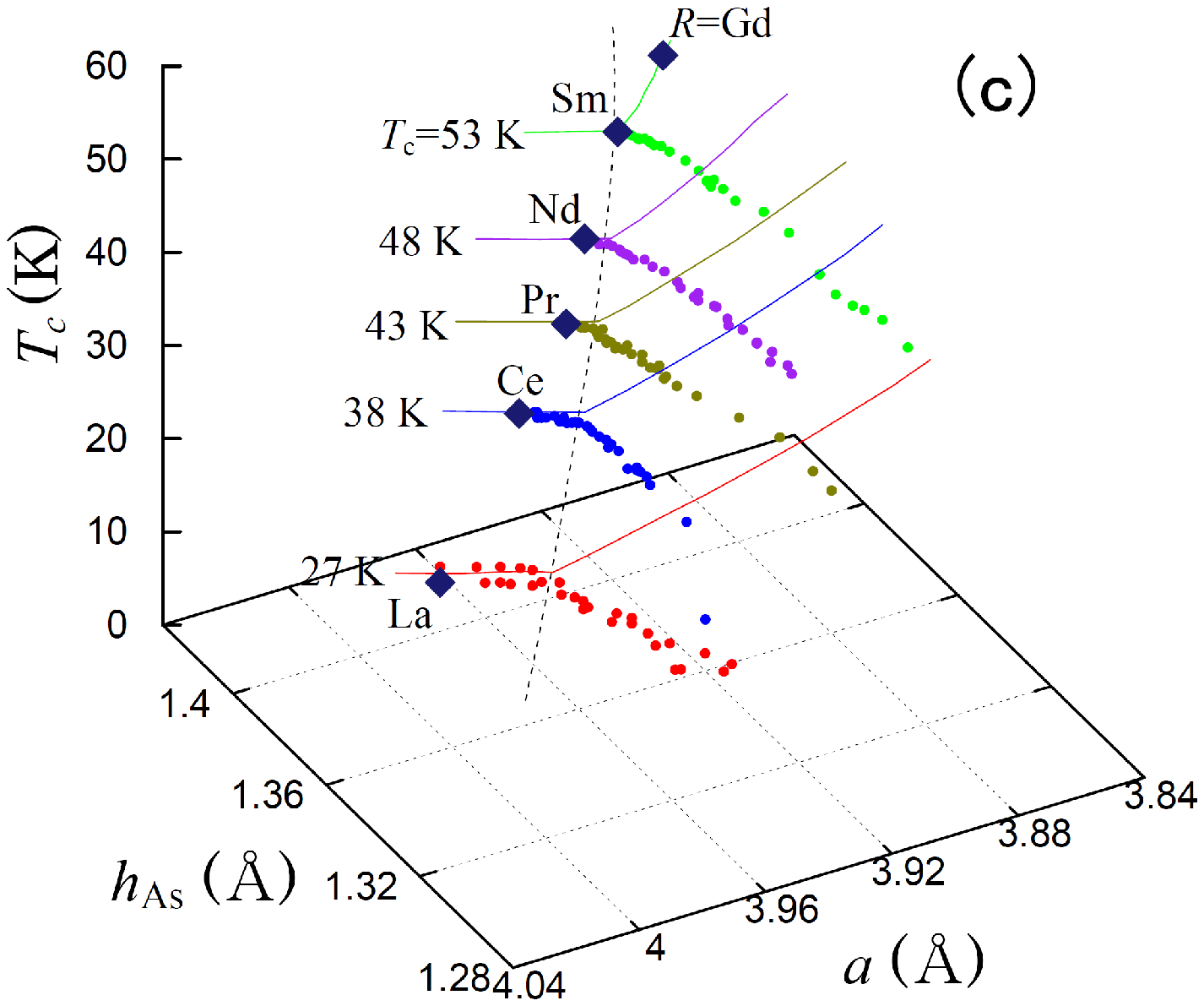}%
\caption{(Color online) $T_{\rm c}$ evolution on the $h_{\rm As}$$-$lattice constant $a$ (a) and 
$h_{\rm As}$$-$$c$ (b) planes for $R$FeAsO$_{\rm 1-x}$F$_{\rm x}$. 
The $T_{\rm c}$($P$) data for each system are plotted on the planes after transforming the data, 
assuming linear relations $h_{\rm As}$$=$$-$3.6$\times$10$^{-3}$$P$$+$$h_{\rm As}$$(0)$, 
$a$=$-$9.3$\times$10$^{-3}$$P$$+$$a$$(0)$ and $c$=$-$4.7$\times$10$^{-2}$$P$$+$$c$$(0)$ 
obtained from the literatures.\cite{garbarino,garbarino2}
The data at ambient pressure for $R$=La, Ce-Gd are displayed by large closed diamonds. 
The solid lines indicate contours of $T_{\rm c}$, which are drawn by connecting the data point with similar $T_{\rm c}$. 
The broken line indicates the upper limit of the lattice constant $a_{\rm ulm}$($h_{\rm As}$) 
($c_{\rm ulm}$($h_{\rm As}$)), above which $dT_{\rm c}$$/$$da$ ($dT_{\rm c}$$/$$dc$) 
changes the sign from positive to negative. 
Three-dimensional plots of $T_{\rm c}$ as a function of $h_{\rm As}$ and $a$ (c). 
The solid lines (broken line) indicate(s) contours (a ridge) of the $T_{\rm c}$($h_{\rm As}$,$a$) surface. 
}
\label{autonum}
\end{figure}

In Fig. 5, we show the plots of $T_{\rm c}$ versus $h_{\rm As}$. 
The $T_{\rm c}$$-$$h_{\rm As}$ curve for the chemical substitution for $R$-site (the broken line) 
is flat in the high $h_{\rm As}$ side, and even in the higher $h_{\rm As}$ region, 
it is flat because $T_{\rm c}$ at ambient pressure 
is known to be 52-53 K for $R$=Sm-Dy.\cite{miyazawa} In contrast, 
$T_{\rm c}$ rapidly decreases in the low $h_{\rm As}$ 
region below the value for $R$=Sm. The $T_{\rm c}$($h_{\rm As}$) 
data for the physical compression for $R$=Sm in Fig. 5 are 
found to be not coincident with the broken line, showing a rapid decrease with decreasing $h_{\rm As}$. 
For $R$=Gd and Tb, $T_{\rm c}$ is known to decrease monotonically with increasing pressure, 
similar to that for $R$=Sm.\cite{takeshita3} 
Since $T_{\rm c}$ is nearly constant for $R$=Sm-Dy at ambient pressure, 
the $T_{\rm c}$($h_{\rm As}$) data under pressure for $R$=Gd and Tb 
are expected to be not coincident with the broken line. 
In addition, a constant part in the $T_{\rm c}$($h_{\rm As}$) data for $R$=La, corresponding to 
the $T_{\rm c}$$-$$P$ plateau, is also not reproduced by the broken line. The discrepancies between 
the $T_{\rm c}$$-$$h_{\rm As}$ relations coming from the physical compression and the chemical substitution 
indicate that $T_{\rm c}$ is determined not only by $h_{\rm As}$ but also by another structural parameter. 

\subsection{$T_{\rm c}$ evolution on $h_{\rm As}$$-$lattice constant plane}
Kuroki $et$ $al$. have pointed that the reduction of the lattice constant $a$ or $c$ suppresses superconductivity 
in $R$FeAsO (i.e., $dT_{\rm c}$$/$$da$$>$0 or $dT_{\rm c}$$/$$dc$$>$0) 
due to the increased hopping integrals and the associated suppression of the electron correlation, and 
therefore the lattice constant is also an important parameter to determine $T_{\rm c}$.\cite{kuroki} 
It should be noted that the lattice constant increases when $h_{\rm As}$ is decreased by the chemical 
substitution but it decreases when $h_{\rm As}$ is decreased by the physical compression, 
changing in opposite directions. 
This could be the origin of the difference in the $T_{\rm c}$$-$$h_{\rm As}$ relations 
derived from the physical compression and the chemical substitution 
if $T_{\rm c}$ also depends on the lattice constant. Kuroki $et$ $al$. have also suggested that 
the effects of $h_{\rm As}$ and lattice constant on $T_{\rm c}$ may cancel 
with each other to result in a nearly constant $T_{\rm c}$ 
between $R$=Nd-Dy at ambient pressure.\cite{kuroki} 
This idea lead us to expect the existence of an upper limit of the lattice 
constant near the value for $R$=Sm, above which $dT_{\rm c}$$/$$da$ (or $dT_{\rm c}$$/$$dc$)
changes the sign from positive to negative, leading to the rapid decrease in $T_{\rm c}$ 
from 53 K ($R$=Sm) to 27 K ($R$=La) due to the combined effect of 
$dT_{\rm c}$$/$$dh_{\rm As}$$>$0 and $dT_{\rm c}$$/$$da$$<$0 ($dT_{\rm c}$$/$$dc$$<$0). 
The plateau in the $T_{\rm c}$$-$$P$ curve observed in the low pressure region 
can be also explained by the existence of the upper limit of the lattice constant. 
When the lattice constant is larger than the limit at low pressure, 
the effect of $dT_{\rm c}$$/$$dh_{\rm As}$$>$0 decreases $T_{\rm c}$ with increasing 
pressure, whereas the effect of $dT_{\rm c}$$/$$da$$<$0 (or $dT_{\rm c}$$/$$dc$$<$0) increases 
$T_{\rm c}$ with increasing pressure. The cancellation of these effects can be the origin of the 
pressure independent behavior of $T_{\rm c}$. 

In order to confirm the importance of $h_{\rm As}$ and lattice constant for 
the superconductivity in $R$FeAsO$_{\rm 1-x}$F$_{\rm x}$, and also the existence of 
the upper limit of the lattice constant, we show $T_{\rm c}$ evolution on the 
$h_{\rm As}$$-$lattice constant plane in Figs. 6(a) and 6(b). 
For the construction of the evolution maps, $T_{\rm c}$($P$) data were 
transformed to $T_{\rm c}$($h_{\rm As}$, $a$) and 
$T_{\rm c}$($h_{\rm As}$, $c$) data, assuming the same linear relations for all compounds, 
$a$=$-$9.3$\times$10$^{-3}$$P$$+$$a$$(0)$ and $c$=$-$4.7$\times$10$^{-2}$$P$$+$$c$$(0)$ 
obtained using the data in the low pressure range ($P$$\leq$8 GPa) in the literature,\cite{garbarino} 
in addition to $h_{\rm As}$$=$$-$3.6$\times$10$^{-3}$$P$$+$$h_{\rm As}$$(0)$,\cite{garbarino2} 
although the relations should be estimated individually using the structural data for each compound 
for the more exact construction. 
Then, $T_{\rm c}$($h_{\rm As}$, $a$ or $c$) data were plotted and 
the contours of $T_{\rm c}$ (solid lines) were drawn by connecting the data point with similar $T_{\rm c}$. 
The data points (small symbols) on the plane represent that both $h_{\rm As}$ and lattice constant for each system decreases 
under pressure from the value at ambient pressure (large closed diamonds). 
The data points above the broken line coincide with the contour of $T_{\rm c}$, 
corresponding to the $T_{\rm c}$$-$$P$ plateau. The contour of $T_{\rm c}$ changes the direction 
across the broken line, indicating that $dT_{\rm c}$$/$$da$ (or $dT_{\rm c}$$/$$dc$) changes the sign 
across the line. Therefore, the broken line corresponds to the upper limit of the lattice constant $a_{\rm ulm}$ 
($c_{\rm ulm}$), which we expect as discussed in the previous paragraph. 
The value of $a_{\rm ulm}$ ($c_{\rm ulm}$) is different for each system, depending on $h_{\rm As}$. 
The qualitative feature of the $T_{\rm c}$ evolution below the $a_{\rm ulm}$ ($c_{\rm ulm}$) line is consistent with that proposed by 
Kuroki $et$ $al$.\cite{kuroki} 
In Fig. 6(c), we show three-dimensional plots of $T_{\rm c}$($h_{\rm As}$, $a$). 
In the figure, $T_{\rm c}$($h_{\rm As}$, $a$) surface is described by contours of $T_{\rm c}$ (solid lines) 
and a ridge of the surface (broken line). 
The evolution map of $T_{\rm c}$ in Fig. 6(a) is a projection of the $T_{\rm c}$($h_{\rm As}$, $a$) surface 
on $h_{\rm As}$$-$$a$ plane, so that the $a_{\rm ulm}$ line is corresponding 
to the projection of the ridge of the surface.  

We have shown in Fig. 5 that the variations of $T_{\rm c}$ derived from the physical 
compression and the chemical substitution can not be expressed 
by a universal function $T_{\rm c}$($h_{\rm As}$). 
On the other hand, the $T_{\rm c}$($h_{\rm As}$, $a$ or $c$) surface 
shown in Figs. 6(a)-6(c) can reasonably describe the characteristic variations of $T_{\rm c}$, i.e., 
the $T_{\rm c}$$-$$P$ plateau followed by the monotonic decrease under pressure 
and the rapid decrease in $T_{\rm c}$ (nearly constant $T_{\rm c}$) 
by changing $R$ elements from Sm to La (from Tb to Sm). 
Furthermore, the $T_{\rm c}$($h_{\rm As}$, $a$ or $c$) surface would 
also describe a monotonic decrease in $T_{\rm c}$ 
under pressure observed for $R$=Gd and Tb.\cite{takeshita3} These facts suggest that 
$h_{\rm As}$ and lattice constant are important structural parameters to determine 
the superconductivity in $R$FeAsO$_{\rm 1-x}$F$_{\rm x}$. 
It is however unclear which of the lattice constant 
is dominant for the superconductivity. The in-plane electron hopping, which is intuitively expected to 
be essential to the superconductivity due to the layered structure, can be decreased by 
the enhancement of either $a$ or $c$.\cite{kuroki} 
The electron transfer in the $c$ direction may be also important in view of the doping from 
the $R$-O$_{\rm 1-x}$F$_{\rm x}$ layer to the superconducting Fe-As layer. 
The dc magnetic measurements under uniaxial pressure using single-crystal specimens are 
necessary to specify the axis sensitive to the superconductivity. 

\section{Summary}
In the present work, we have performed the dc magnetization measurements under pressure for 
optimally doped $R$FeAsO$_{\rm 1-x}$F$_{\rm x}$ ($R$=La and Ce-Sm). 
It is found that the $T_{\rm c}$$-$$P$ curve for $R$=La exhibits 
a plateau at low pressure range up to $\sim$3 GPa, followed by a monotonic decrease at higher pressure. 
The plateau width in the $T_{\rm c}$$-$$P$ curve is found to depend on the lattice constant of the system and shrinks to 
nearly zero for $R$=Sm. Although $h_{\rm As}$ is known to be 
an important structural parameter, the variations of $T_{\rm c}$ derived from the 
physical compression and chemical substitution for $R$-site can not be expressed 
by a universal function $T_{\rm c}$($h_{\rm As}$). Instead, we present the $T_{\rm c}$ evolution map on 
the $h_{\rm As}$-lattice constant plane, where it is shown that 
$T_{\rm c}$($h_{\rm As}$, $a$ or $c$) surface can describe all of the characteristic 
$T_{\rm c}$ variations, suggesting that $T_{\rm c}$ is determined by 
$h_{\rm As}$ and lattice constant. In addition, the upper limit of the 
lattice constant $a_{\rm ulm}$ or $c_{\rm ulm}$, across which $dT_{\rm c}$$/$$da$ (or $dT_{\rm c}$$/$$dc$)  
changes the sign, has been shown to exist.  
The $T_{\rm c}$$-$$P$ plateau observed for $R$=La and Ce-Nd is thought to be originating from the effects of $h_{\rm As}$ 
and lattice constant on $T_{\rm c}$ canceling each other for $a$$>$$a_{\rm ulm}$ ($c$$>$$c_{\rm ulm}$).

\begin{acknowledgments}
This work is financially supported in part by a Grant-in-Aid for 
Scientific Research (No. 20540355) from the Japanese Ministry of Education, 
Culture, Sports, Science and Technology. 
\end{acknowledgments}


\end{document}